\documentclass[12pt,a4paper,final]{iopart}

\usepackage{iopams}  
\usepackage{graphicx}
\usepackage[breaklinks=true,colorlinks=true,linkcolor=blue,urlcolor=blue,citecolor=blue]{hyperref}
\usepackage{cite}
\usepackage{longtable}
\begin{document}

\title[Plasma Sources Science and Technology]{Strong Evidence of Plasma Like Behavior for Ion-Solid Collisions}

\author{Prashant Sharma}
\address{Inter University Accelerator Center, Aruna Asaf Ali Marg, Delhi 110067}
\ead{prashant@iuac.res.in}

\author{Tapan Nandi}
\address{Inter University Accelerator Center, Aruna Asaf Ali Marg, Delhi 110067}
\ead{nandi@iuac.res.in}

\begin{abstract}
Charge state distributions of various projectile ions passing through thin carbon foil have been studied in the energy range of 0.7-3.0 MeV/u using x-ray spectroscopy. This technique is found to be appropriate to segregate the charge state distribution in the bulk to a great extent from that of the surface by measuring the charge changing phenomena right at the ion-atom interaction zone. This observation has been confirmed by different theoretical approaches. Surprisingly, it is found that the charge state distribution measured in the bulk, exhibits Lorentzian profile which is an important characteristic of any plasma. The occurrence of such behavior suggests that ion-solid collisions constitute tenuous high density plasma in the bulk of the solid target. This beam-foil plasma is similar to high density stellar interior plasma, which may have practical implications in various fields, in particular, plasma physics and astrophysics. One important point is to note that the theoretical calculations display a Gaussian structure for the charge state distribution. Thus, this work suggests the inclusion of plasma coupling effects in theoretical calculations , which is not yet taken into account in the theory, to explain the distribution observed.

\end{abstract}

\pacs{34.50.Fa, 
34.70.+e,  
52.20.Hv, 
32.70.Jz 
} 
\vspace{2pc}
\submitto{\JPB}

\section{Introduction}

Atomic phenomena such as electron-capture and -loss processes \cite{mac,thor,shev} are the key factors for any collisions of swift ions with atoms, ions or molecules that cause a change in charge state of the ions traversing a medium. Even though a monochromatic ion beam with a fixed charge state is passed through the medium, several charge states emerge out of the target irrespective of its thickness \cite{shima}. However, after a large number of collisions, equilibrium in charge state distribution (CSD) as well as mean charge state (q$_{m}$) is established, when certain balance in electron-capture and -loss processes is attained. The number of collisions or thickness required to arrive at the equilibrium depends on the ion species, its velocity and characteristics of the target including atomic number, density, phase, structure etc.
	Experimental techniques involved in measuring CSD and q$_{m}$ are mainly electromagnetic in origin, thereby accounting for the total charge of the ion in the detectors placed at the focal plane, a few meters away from the target (time of flight is of the order of $\mu$sec). This implies that these techniques give an integral measure of electron-capture and -loss processes at bulk as well as surface of the foil and cannot allow one to segregate the charge changing phenomenon in the bulk from that of the surface. This difficulty can be circumvented using charge less observables in the experiments. In the present work, we intend to study the ion-solid interactions only in the bulk. With this motivation, we confine the work to study the CSD and other relevant parameters right at the interaction zone using the x-ray spectroscopy technique.

\section{Experiment}
Experiments were performed with the energetic ion beams of $^{58}$Ni and $^{56}$Fe using 15 UD Pelletron \cite{Kanjilal} accelerator at IUAC, New Delhi. Well-collimated ion beam of energies 0.7-3.0 MeV/u starting from deep sub-barrier to above the barrier energies were bombarded on 80 $\mu$g/cm$^{2}$ thick amorphous carbon (natural) target foils. The foil thickness was chosen such that even the highest beam energy used attains the equilibrium charge distribution. The target was placed at 45$^{\circ}$ to the beam axis so that we can measure the x-ray spectra right from the ion-solid interaction zone. The x-ray produced in the ion-solid interactions passes through two collimator of 5 mm diameter kept at 5 cm apart and the first collimator is placed at 65 cm away from the target. It thus ensures the x-rays coming from a tiny section ($\pm$1 mm) of the interaction zone. In the time scale the detector sees only about $\pm$50 psec with respect to the center of the interaction zone. Hence, the x-ray spectroscopy can be considered as a measurement at t=0 compared to the electromagnetic measurements taking place at t=t', t' $\approx$ a few $\mu$sec. The x-rays were detected in a Low Energy Germanium Detector (GUL0035, Canberra Inc., with 25 $\mu$m thick Be entrance window, resolution 150 eV at 5.9 keV, with constant quantum efficiency in the range of 3-20 keV) at 90$^{\circ}$ to the beam axis. At this geometry Doppler broadening is maximum, in contrast the Doppler shift is minimum, $E_{0}$[1-$\gamma^{-1}$] (the beam energies used give rise to $\gamma$ $\sim$ 1, hence shift tends to zero). The detector was kept outside the chamber through a thin mylar window of 6 $\mu$m at the interface. The detector observes both the ion-entrance as well as the ion-exit side of the ion-solid interaction. The beam was dumped in a Faraday cage. Two solid surface barrier detectors were used at $\pm$10$^{\circ}$ to monitor the beam particles. Vacuum chamber was maintained at a pressure around 1$\times$10$^{-6}$ Torr. The x-ray spectra observed for all the beam energies are shown in Fig. \ref{fig1}. Calibrations were done for the x-ray detectors using $^{60}$Co and $^{241}$Am standard sources. The resolution was found to be about 200 eV at 6.4 keV with the experimental conditions in the beam-hall.

\section{Data Analysis, Results and Discussions}

The primary motivation of this work is to determine the charge state distribution of projectile ions at the interaction zone along with q$_{m}$ from the measured x-ray spectra. It is important to note that the parameters obtained at different beam energies for the particular ion can be compared without normalizing the x-ray spectra. Hence, complexity of normalization is avoided in this work like any electromagnetic method coupled with position sensitive detectors \cite{Stohlker}.
 \begin{figure*}[!ht]

\includegraphics[scale=1.5]{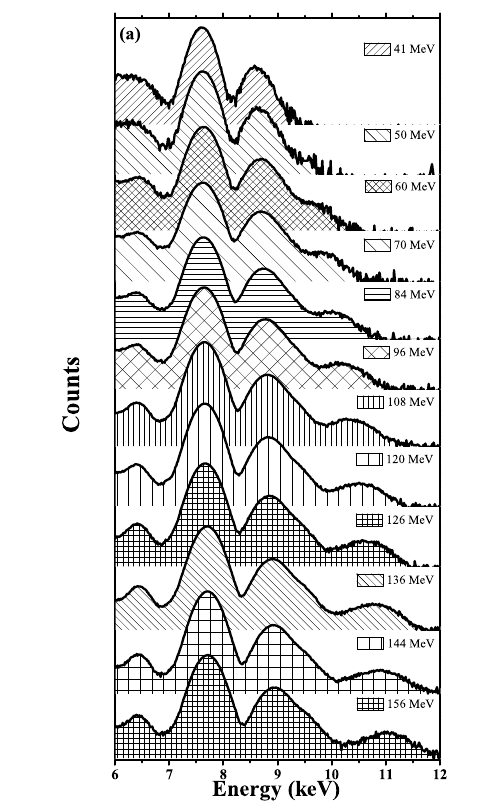}
\includegraphics[scale=1.5]{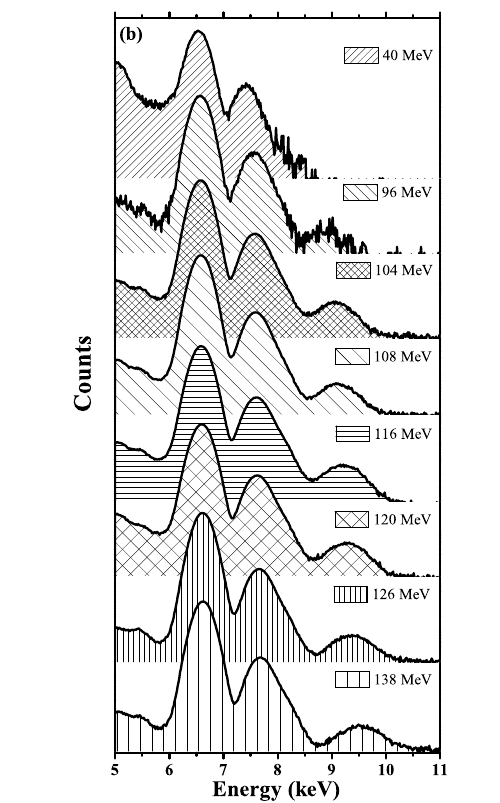}

\caption{\label{fig1} X-ray spectra for (a) $^{58}$Ni beam and (b) $^{56}$Fe beam on 80 $\mu$g/cm$^{2}$ C-foil at different beam energies 
}
\end{figure*}
 \begin{figure}[!ht]

\includegraphics[scale=.5]{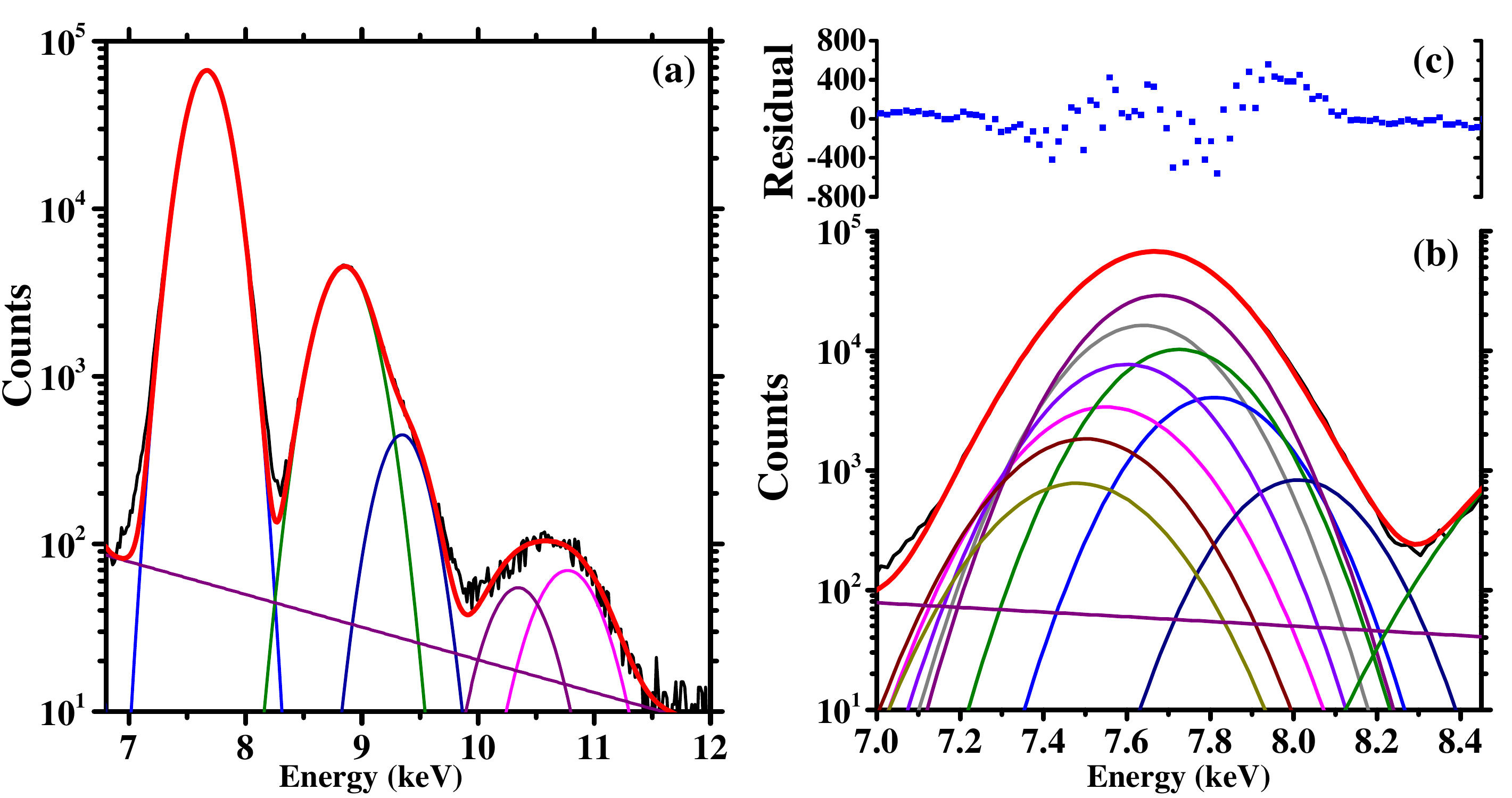}

\caption{\label{fig2} X-ray spectrum of $^{58}$Ni on C at 126 MeV (a) Fitting shows only broad features of the spectrum (b) Projectile x-ray peak is fitted into nine Gaussian functions corresponding to x-ray lines appearing from H-like to F-like Ni with a exponential function representing the background (c) The residuals of fitting (b) is shown }

\end{figure}
\noindent The spectrum displays four peak structures at 6.4, 7.7, 8.8 and 10-12 keV, the case of 126 MeV Ni on C as shown in Fig. \ref{fig1} and Fig. \ref{fig2}. The structure at 6.4 keV appears due to beam halo falling on the target frame made of stainless steel. This gives us an opportunity to calibrate internally the spectra while Ni ion-beam is used. However, in case of Fe-beam experiment, beam halo was minimized by passing the beam through a blank target frame so that its presence does not affect much the peak structure originated from the projectile ions. The peak structure of 7.7 keV represents the x-ray peak from the projectile ions, whereas structure at 8.8 and 10-12 keV are due to $\alpha$ and 2$\alpha$ capture events, respectively \cite{Nandi}. It is worth mentioning here that the third and fourth structures are of no relevance in this work, hence they will not appear in further discussion. From the Fig. \ref{fig1} it is discernible that as the beam energy increases, centroid of the projectile x-ray peak keeps on shifting towards higher energy side. 
 
 When a fast projectile ion bombards on any target atom, the interaction gives rise to vacancies in the different shells of the target atoms, which causes energy shift of the characteristic x-ray lines and production of hyper-satellite lines \cite{ajay}. In contrast, such phenomena do not occur with the projectile ions as most of the bound outer electrons are stripped off during the ion-solid interactions \cite{Schnopper}. Hence the x-ray lines belonging to the projectile ions are attributed to the transitions due to a single vacancy \cite{ajay}. Further, for every charge state there is a certain probability of creating a single K-shell vacancy owing to ion impact on the target with a certain energy that gives rise to a characteristic x-ray emission for that particular charge state. Therefore, we expect many x-ray lines emanating from different charge states of the projectile ions. Further the x-ray spectroscopy is used as a reliable method to find the charge state distribution in various plasma  \cite{Santos,Ullmann,Friedlein,Guerra,Rosler} and therefore in this work we plan to utilize the technique for the ion-atom collision. Though detector resolution in the experiment restricts us to resolve individual x-ray lines; the well-defined centroid gives a correct measure of the mean charge state. 

 In order to get a right correspondence between the centroid and the mean charge state, we have adopted a special analysis method as follows: in first step, we plot the standard K$_{\alpha}$ x-ray energies from NIST database \cite{Ralchenko} for H-like to F-like Fe ions. Next, in order to know the correct q for any x-ray energy, we have fitted the data with a multi-parametric equation as a function of the x-ray energies. Afterwards, projectile x-ray peak is fitted with a Gaussian function to find the centroid and corresponding charge state is computed from the multi-parametric equation discussed above. The charge state so obtained represents the value of q$_{m}$. It should be noted here that K$_{\alpha}$ x-ray energies for Ni are available only for H-like to Li-like Ni. The rest of the energies for Be-like to F-like Ni are scaled from corresponding Fe data \cite{Ralchenko}. This procedure is repeated for both the ion species of all beam energies. 
  \begin{figure*}[!ht]

\includegraphics[scale=.3]{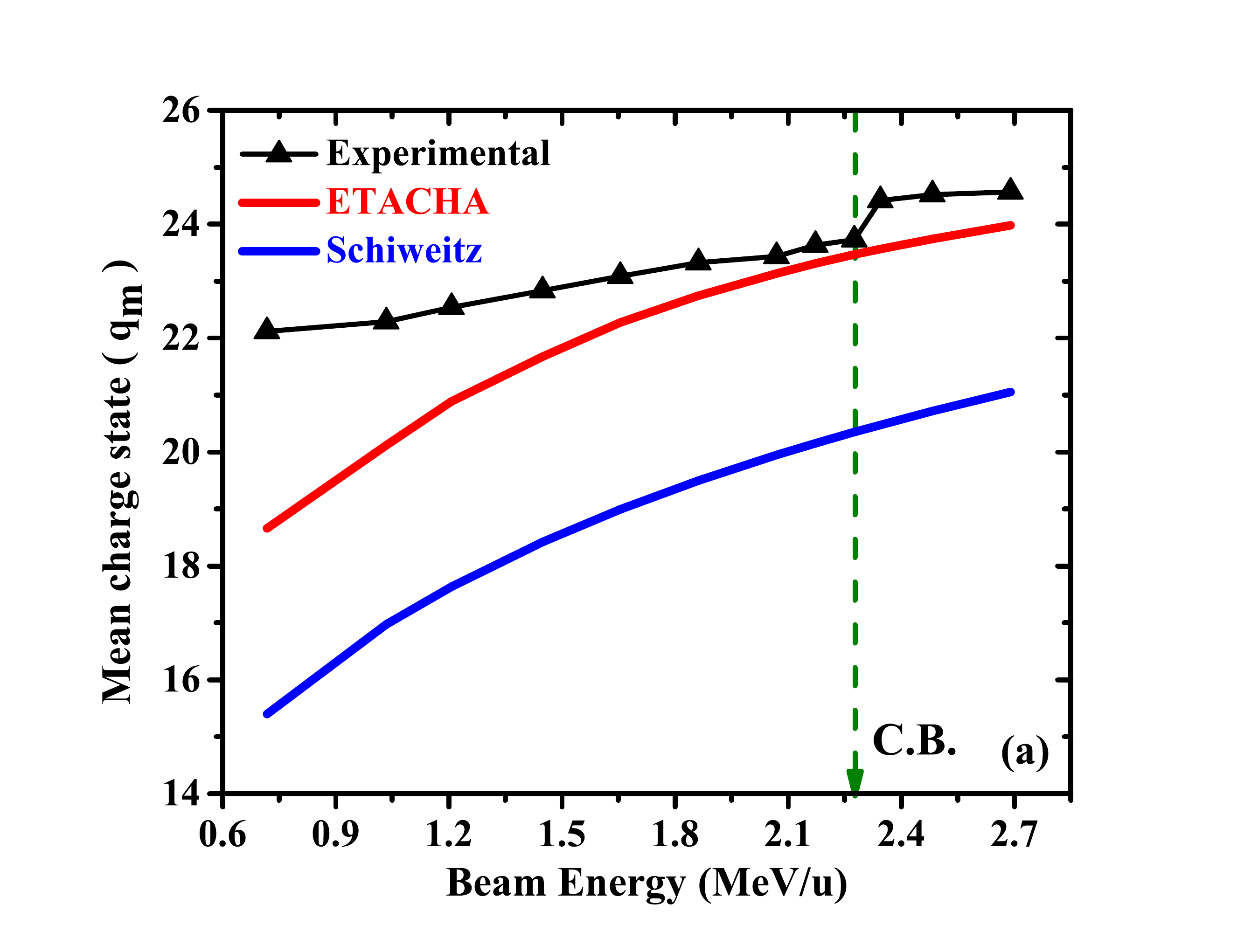}
\includegraphics[scale=.3]{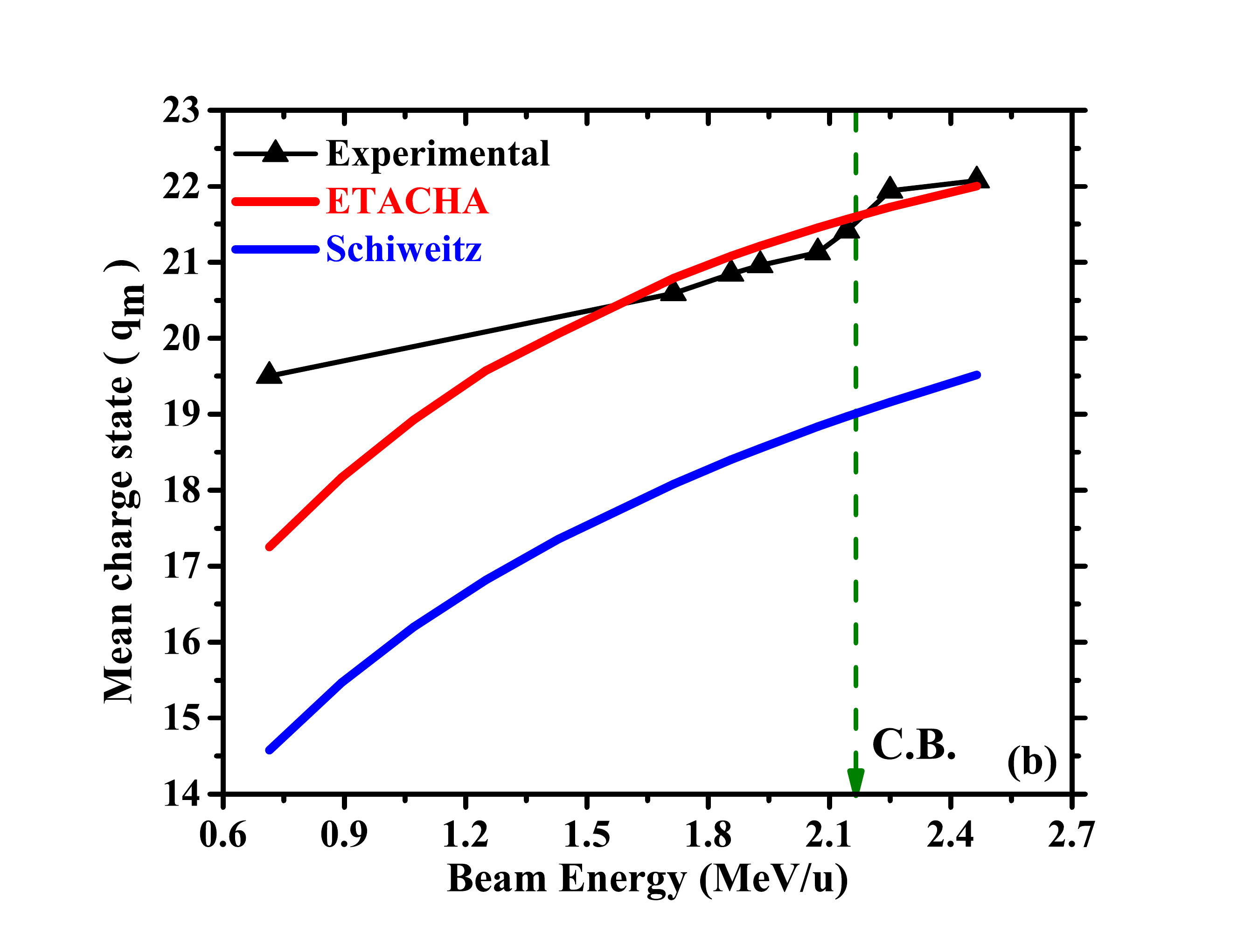}
\caption{\label{fig3} Mean charge states versus the beam energies for (a) $^{58}$Ni beam and (b) $^{56}$Fe beam on 80 $\mu$g/cm$^{2}$ C-foil. Solid lines are to guide eye only. Error bars are smeared with the symbol size.
}
\end{figure*}	
 In next step, for validating that our measurements represent bulk effect only, we choose two different approaches for comparison. First, ETACHA code \cite{Rozet} which takes account of ionization and capture processes theoretically and ought to represent the measurements in bulk, is used to get q$_{m}$ for Ni and Fe ions on 113 $\mu$g/cm$^{2}$ thick C target at different beam energies. It is important to note that empirical formalisms, which are tuned on the basis of the measured data taken from the electromagnetic methods, represent the integral role of the bulk and the surface of the foil. Thus, in the midst of available empirical formalisms we choose Schiwietz formalism \cite{Schiwietz} because of its vast and updated dataset \cite{Schmitt} to calculate q$_{m}$ for a combined role of bulk and surface effect. The experimentally measured values of q$_{m}$ are compared with ETACHA predictions and Schiwietz formalism as shown in Fig. \ref{fig3}. As expected, Schiwietz calculations \cite{Schiwietz} are found to be much lower than both ETACHA predictions \cite{Rozet} and present experimental results, showing a clear indication of dominant multi-electron capture from the surface of the foil \cite{Brauning}. In contrast, ETACHA predictions show quite good agreement with experimentally measured values of q$_{m}$. However, for Ni case, experimental q$_{m}$ are little greater than the ETACHA predicted values for $<$ 2MeV/u. Whereas for Fe case, experimentally measured q$_{m}$ are found little lesser in the same energy range. Interestingly, for both cases q$_{m}$ values start merging at $\geq$ 2 MeV/u. Thus, the comparison not only fairly validates the surface effect removal in the x-ray spectroscopy technique, but also reveals that ETACHA predictions \cite{Rozet} better represent the data obtained at t=0 and approached to the measured data for energies $\geq$ 2 MeV/u. Hence, our work also establishes that ETACHA \cite{Rozet} does not only represent data beyond 10 MeV/u as one of our earlier studies revealed \cite{Ahmad}, but also the lower energy side too. Surprisingly, at the low beam energies ($\sim$0.6-1.0 MeV/u); ETACHA predictions underestimate the experimental observations in case of both Ni and Fe. Ionization accompanying electron shake processes due to inner shell ionization \cite{Mukoyama} caused by Auger decay and other autoionizing transitions \cite{Rudek,Young} can probably be responsible for such happening. These processes are more prominent with low charged ions. That is not considered in the ETACHA calculations. In addition to this, one can notice that in both cases, around the Coulomb barrier (CB) unusual enhancement in ionization is observed, which results a kink in experimentally measured q$_{m}$, as shown in Fig. \ref{fig3}. The CB energy in lab frame for $^{58}$Ni and $^{56}$Fe on $^{12}$C are 2.313 and 2.167 MeV/u respectively \cite{Christensen}. Exploring the possible reason of such occurrence is beyond the scope of this paper.

	It is noteworthy to mention that there are certain atomic processes like formation of long lived Rydberg states occurring at the exit surface of the foil \cite{Nandi2,Schiwietz2,Day,Wilhelm} which can introduce another CSD on each charge state produced in bulk of the foil. Thus in the next step, we have attempted to measure the CSD in bulk for each beam energy by limiting the measurements right at the ion-atom collisions. Since, low resolution of solid state x-ray detector used in this experiment gives a broad single peak accounting for the x-ray emissions from projectile ions of several charge states; we have adopted a method to de-convolute individual x-ray intensities corresponding to each charge state. Each peak has been fitted with many Gaussian functions by fixing the centroid of the peaks to H-like to F-like ion x-ray and an exponential function as the background as shown in Fig. \ref{fig2}b. One can note that number of allowed transitions increases with number of electrons available to fill in the K vacancy. As a result the energy level diagram becomes more and more complex as it goes from higher to lower ionic state. Hence, full width at half maximum of the K$_{\alpha}$ x-ray lines, were fixed in ascending order from H-like to F-like ions.

One can further see that Fig. \ref{fig2}a depicts only the broad features of the spectrum. Whereas Fig. \ref{fig2}b displays the feature like intensity contribution from each ionic state of H-like Ni to F-like Ni x-rays. The residuals of the fitting shown in Fig. \ref{fig2}c validate the good quality of fitting performed. In this way we have obtained the distribution of charge state fractions (F$_{q}$) directly from the measured distribution of intensities as follows

 \begin{equation}  
 F_{q} = \frac{I_{q}}{\sum\limits_{q} I_{q}}
 \label{Lsrpulse} 
 \end{equation}
 \begin{figure}[!hbp]

\includegraphics[scale=.4]{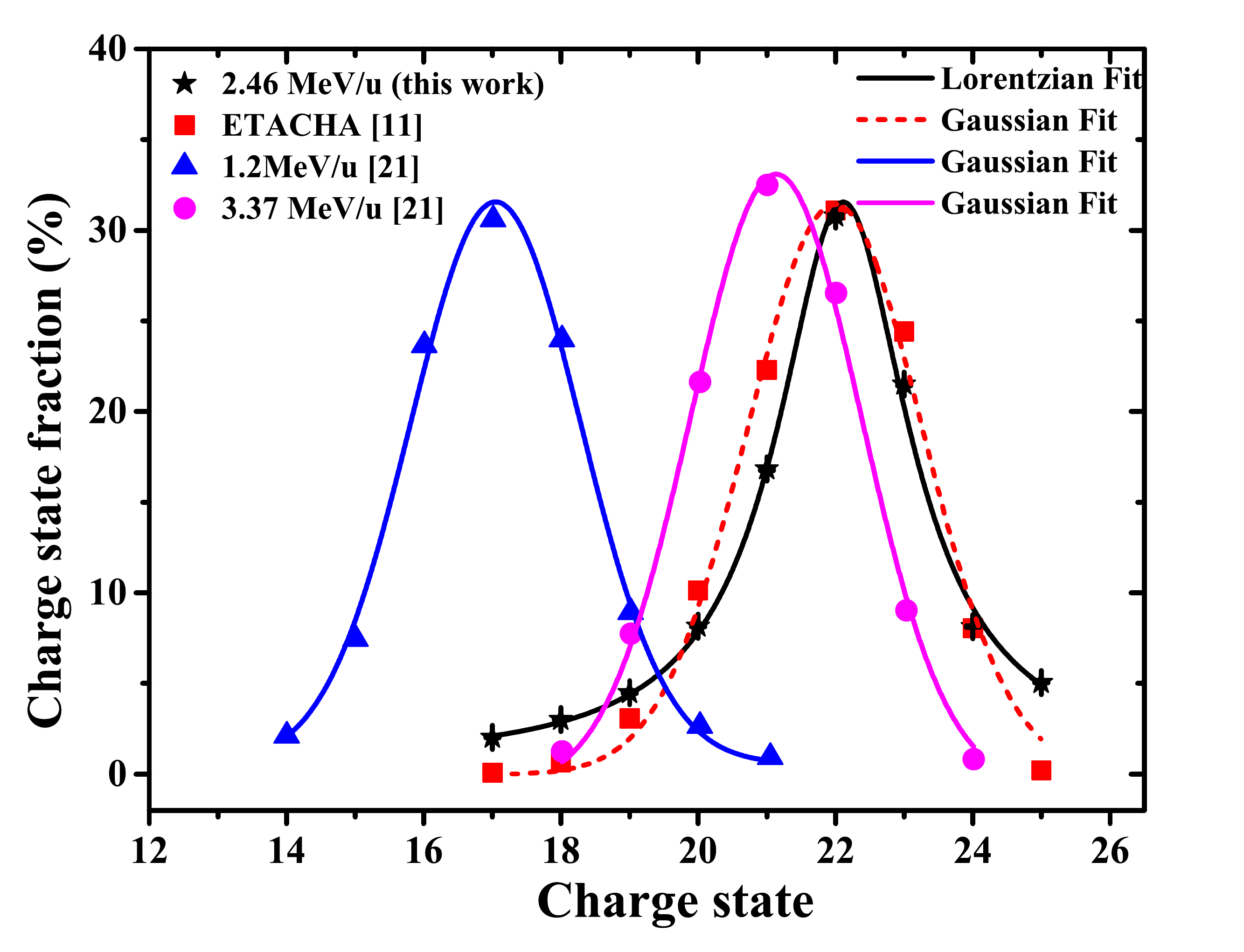}

\caption{\label{fig4} Comparison of the CSDs from reported experimental data using electromagnetic method (t=t') \cite{McMahan}, ETACHA predictions \cite{Ralchenko} and the present data with x-ray spectroscopy method (t=0) for $^{56}$Fe for different beam energies. Figure shows Lorentzian fit to present work and Gaussian fit to ETACHA \cite{Ralchenko} and experimental work \cite{McMahan}. Errors are embedded in symbol itself.
}
\end{figure}  Here, I$_{q}$ is the intensity corresponding to the charge state q. To include other sources of errors besides the statistical error, we have added an error equal to two times the statistical error in every F$_{q}$. Two representative CSD using electromagnetic techniques \cite{McMahan} along with a typical CSD from this work and ETACHA predicted values are displayed in Fig. \ref{fig4}. McMahan et al. have measured the CSD with 1.2 and 3.37 MeV/u, whereas we did with 2.46 MeV/u $^{56}$Fe ion beam. One expects the current measurement must fall in between the two. But it falls even beyond the 3.37 MeV/u data. This picture reveals once more the effect of the surface as discussed earlier.
  \begin{figure*}[!ht]

\includegraphics[scale=.85]{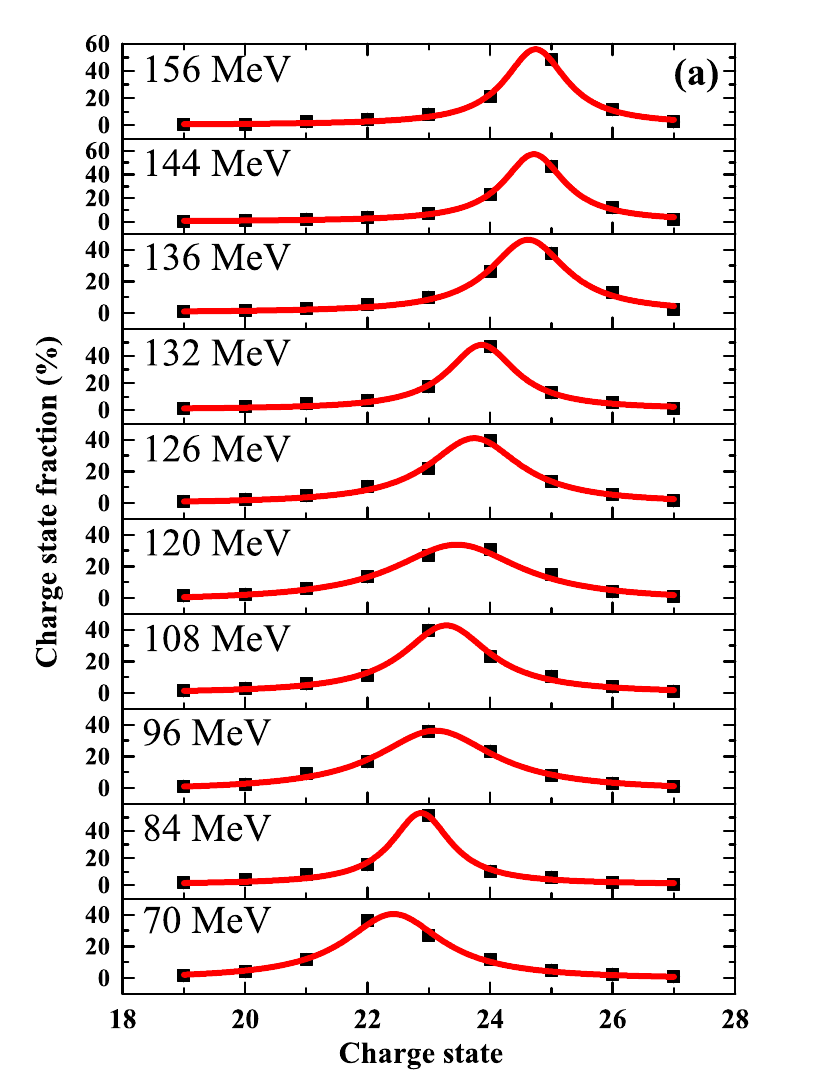}
\hfill
\includegraphics[scale=.85]{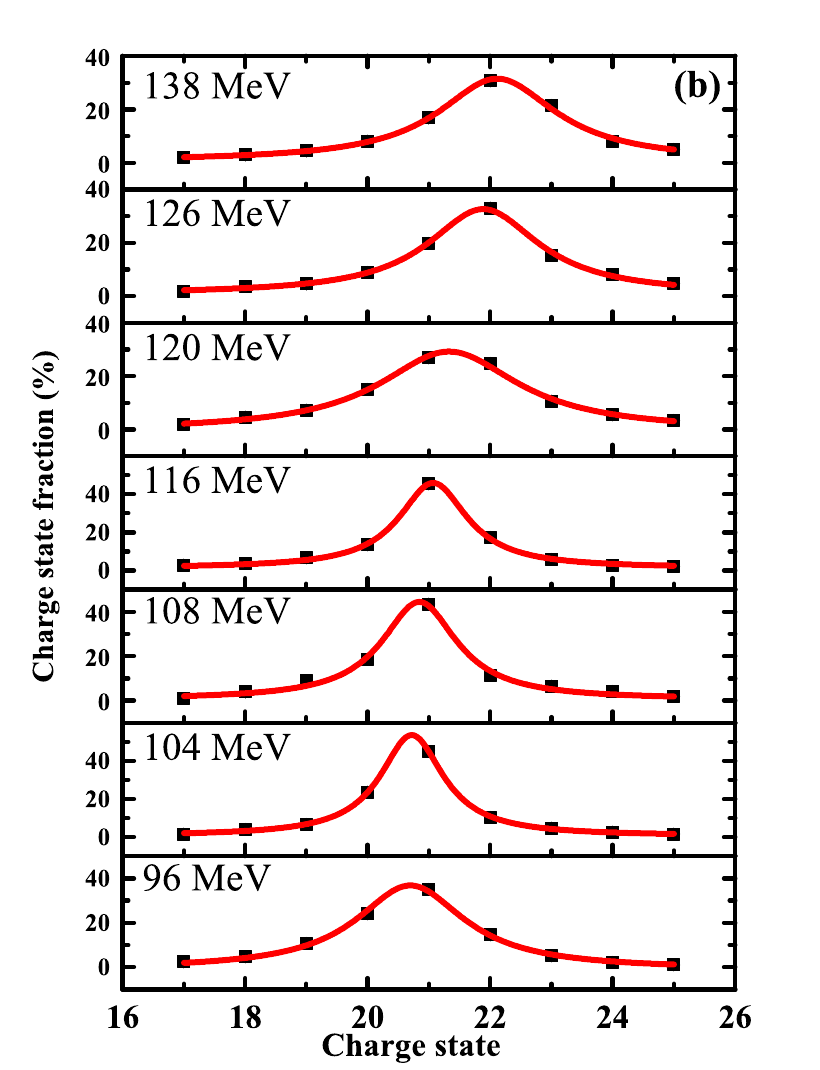}

\caption{\label{fig5} Measured CSD with Lorentzian fitting for (a) $^{58}$Ni on C and (b) $^{56}$Fe on C for various beam energies.
}
\end{figure*}
  The CSD predicted by ETACHA as well as the measured data at t=t' follows the Gaussian shape, however, present dataset depicts a different pattern, which is fitted well with a Lorentzian function for all the cases, shown in Fig. \ref{fig4} and \ref{fig5}. It may be worth noting here that, deviation from the Gaussian distribution can be observed in various plasmas, for example, Voigt distribution has been found in Ar plasma produced in electron cyclotron resonance ion source \cite{Guerra} and Lorentzian distribution for z-pinch plasma \cite{Foord}. Further, CSD of gold ions in EBIT plasma shows also Lorentzian distribution \cite{Wong} that has also been theoretically reproduced \cite{Peyrusse}. Further, besides the laboratory plasma, astrophysical plasma exhibiting the same distribution is popularly known as Lorentzian astrophysical plasma \cite{Jiang,Jung,Ki}. Thus this work reveals that interactions due to ion beam passing through solid thin foil replicate the typical plasma behaviour. Such plasma can be considered as beam-foil plasma having very high density, but extremely small volume. Tenuous plasma is also produced during laser-solid interaction \cite{Von} in the laboratory. Similar high density plasma is prevalent in stellar interiors \cite{Gill} in large volume.  The plasma exhibits long- and short-range order due to the correlating effects of the atoms and ions. The charge state distribution calculations such as ETACHA have not yet taken, the plasma coupling effects into consideration. This may be the possible reason on the difference observed in the charge state distribution between the present measurement and the ETACHA predictions as discussed above. Thus, this work can lead to provide useful informations to understand high density plasma where atoms and ions behave in a manner intrinsically coupled to the plasma itself.
  
We must mention here that the high density plasma by ion-solid interactions were found to be produced with intense heavy ion beams of the order of 10$^{11}$ ions in 1$\mu$ sec pulse with high energy beams 300 MeV/u in GSI Darmstadt \cite{Dewald}. In contrast we observed the high density plasma by the impact of 10$^{10}$ particles/sec DC ion beam with 0.7-3.0 MeV/u heavy ions on thin C-target, revealing that the low beam current and few MeV/u energy also can give rise to high density plasma in solid targets. Further, it will be interesting to check the ranges of beam conditions that can give rise to high density plasma in solid targets.
\section{Conclusion}
The charge state distribution of the projectile ions during ion-solid collisions can only be carried out at t=0 by charge-less observables, as is done here using x-ray spectroscopy. This approach succeeds in observing the CSD only at the bulk of the foil unlike the electromagnetic techniques, which give an integrated contribution from both the bulk as well as the exit surface. Difference between the measured q$_{m}$ in this measurement and the calculated CSD from empirical formalism suggests multi-electron capture from the exit surface \cite{Brauning}. Further, it is shown that ETACHA code \cite{Rozet} represents well the measurements  at t=0 for energies $\geq$ 2 MeV/u. Though an unusual charge state distribution in the form of Lorentzian distribution is observed in contrast to the ETACHA predictions of the Gaussian distribution. Appearance of Lorentzian distribution for the CSD is analogous to the CSD in any plasma.  The difference found between the observed and ETACHA predicted CSD may be attributed to the plasma coupling effects that is not considered in the theory. Thus, ion-solid collisions in the bulk of a solid target may represent well a major characteristic of the plasma. Special property inheriting high density plasma in the laboratory is rare to simulate the characteristics of high density plasma as seen in stellar interior \cite{Gill}. We believe the studies using such plasma will open up new opportunities to the researchers.

\section*{Acknowledgement}

 We would like to acknowledge the co-operation and support received from the Pelletron accelerator staff and all colleagues of Atomic physics group, IUAC, New Delhi. PS is thankful to UGC, India for providing the fellowship as financial support to carry out this work.

\end{document}